\documentclass[runningheads]{llncs}
\usepackage{microtype}

\usepackage{graphicx}
\usepackage{color}
\usepackage{amsmath,amssymb,amsfonts}
\usepackage{xcolor}
\usepackage{hyperref}
\usepackage{booktabs}
\usepackage{cite}

\begin{document}

\title{Thresholding Post-Quantum Signatures}
\author{
Francesco {De Sclavis}\inst{1,2}
\and
Matteo Nardelli\inst{1}
\and
Marco Pedicini\inst{2} 
}
%
%
\institute{Bank of Italy\\
\email{\{firstname.lastname\}@bancaditalia.it}
\and
Roma Tre University, Rome, Italy \\
\email{marco.pedicini@uniroma3.it}
}
\maketitle

\begin{abstract}
Threshold signature schemes distribute the signing process among $T$ parties out of $N$. They enable a variety of applications and their research is also motivated by a recent NIST call. However, applications are dominated by pre-quantum signatures, which are more efficient but not secure in the post-quantum setting. This paper investigates existing post-quantum signatures, based on a variety of paradigms: lattice problems, one-way (hash) functions, cryptographic group actions, isogenies and multivariate systems. We propose a classification (divided by paradigm) of existing tools that are used to build $T$-out-of-$N$ schemes from digital signatures. We also include general approaches based on FHE, MPC or ZKP.
\end{abstract}

\section{Introduction}

A threshold signature is a cryptographic primitive enabling a group of $T$ signers out of a total of $N$ participants to jointly produce a valid signature, while preventing $T-1$ dishonest participants from forging it. The basic idea is to split the secret key into partial secrets, which are used to produce partial signatures that are later aggregated. The benefit of this method is to distribute trust across several users and prevent single-points of failure. Recently, threshold signatures have been employed for several applications (e.g.,~\cite{applicationConsensus,applicationThresholdECDSA,applicationDistributedSSH,applicationDNSSEC,applicationCandid,applicationLNGate,applicationSmartContract,applicationTrustSAS}), such as consensus for blockchains or distributed key management. Interest in threshold cryptography has been also motivated by institutions, such as NIST, that has recently published a call for standardization\footnote{\url{https://csrc.nist.gov/projects/threshold-cryptography}}.

A variety of efficient threshold schemes exists---based on RSA, DSA, ECDSA, BLS or Schnorr~\cite{thresholdSurvey}. However, there are few post-quantum (PQ) solutions, which are often less practical due to the size of keys or signatures, or signing times. The threat of quantum computers has motivated research for cryptography based on quantum-safe assumptions and, in recent years, NIST has been conducting a competition to select and standardize PQ schemes\footnote{\url{https://csrc.nist.gov/projects/post-quantum-cryptography}}. Unfortunately, the standardization process is only concerned with simple signatures and not threshold ones. We also observe that the schemes proposed in the competition are not immediately convertible to threshold versions using MPC~\cite{CozzoSmart19}.

\paragraph{Techniques for threshold PQC}
Techniques used for PQ threshold signatures are often borrowed from previous classical works. However, they have to be adapted to take into account the specific problems that arise in the PQ setting. 
%
Most of the techniques employ some form of \textit{linear secret sharing} and follow an approach similar to Schnorr-like threshold signatures, such as FROST~\cite{KG2020}, by leveraging the linearity of the underlying scheme. Most of the effort involves adapting this template to specific settings, such as dealing with non-linear operations in lattice-based schemes, or using group actions, which generalize discrete logarithm constructions. On the other hand, hash-based signatures are completely non-linear and have to resort to their own techniques. More general approaches employ other cryptographic primitives: \textit{threshold fully homomorphic encryption (TFHE)} or \textit{multi-party computation (MPC)} to securely evaluate the signature; \textit{zero-knowledge proofs (ZKP)} to prove that aggregated signatures verify correctly\footnote{This is only used by one hash-based scheme~\cite{Khaburzaniya2022}.}. We remark that FHE and MPC are very general techniques that could, in principle, be applied as black boxes, but doing so is often impractical, unless the underlying scheme is FHE- or MPC-friendly. The same is true for ZKPs, that additionally require an algebraic representation of the verification process.

\paragraph{Contribution and outline}
We investigate existing PQ threshold signatures and examine the techniques with which they are implemented. We aim to provide a reference for future research towards an approach that is as modular as possible, by presenting a systematization of existing tools employed in the different PQ paradigms. We focus on individual techniques, rather than individual threshold signatures, while existing works either are comprehensive surveys without a specific focus on PQ \cite{thresholdSurvey, thresholdSurvey2025} or investigate only a class of techniques \cite{CozzoSmart19}.
We are only concerned with $T$-out-of-$N$ schemes, and do not include multi-signatures or ring signatures. In Section~\ref{latticeBased} we examine the techniques used for lattice-based signatures, in Section~\ref{otherParadigms} we explore those used in other paradigms (i.e., hash-based and group-action-based schemes). In Section~\ref{genericApproaches}, we describe some of the ``generic approaches'' that can be applied to any paradigm.

\section{Lattice-based Signatures} \label{latticeBased}
Existing lattice-based signatures are typically based on either the \textit{hash-and-sign} paradigm~\cite{BR1996} or the \textit{Fiat-Shamir with Aborts} (FSwA) paradigm~\cite{Lyu2009}. Some of the operations required by either of the two paradigms are non-trivial to perform in a threshold setting. Therefore, most of the work on lattice-based threshold signatures focuses on how to extend or bypass them entirely. Security of the schemes is based on Short Integer Solution (SIS), Learning with Errors (LWE), or their variants.

The idea of the hash-and-sign paradigm is the following: in the random oracle model, it computes the hash of the message, and then computes the signature as a short pre-image through a trapdoor function (i.e., a function that is hard to invert without having special secret information). It is usually instantiated through the GPV framework~\cite{GPV2008}, that uses a \textit{trapdoor sampler}, i.e., an algorithm that uses the trapdoor to sample the pre-image. This part is usually hard to do in a threshold setting.

FSwA is based on ideas from Lyubashevsky's signature~\cite{Lyu2009,Lyu2012}. Instead of using trapdoors, it constructs an identification scheme from a \textit{sigma protocol}, then converts it into a non-interactive signature by applying the Fiat-Shamir heuristic~\cite{FS1987}. However, this adaptation is not straightforward, because lattice-based problems require short solutions. This means that constraints must be checked to ensure that the solution is valid, and otherwise the scheme is aborted (\textit{rejection sampling}). However, aborting in a threshold setting is not trivial because of possible information leak.
\begin{remark}
    In this section, we assume that the response is $\mathbf{z}=\mathbf{r}+c\cdot\mathbf{s}$, where $\mathbf{s}$ is the secret key and $\mathbf{r}$ is a random element. The notation is from FSwA, but the GPV framework (in particular with the \textit{gadget matrix} technique~\cite{MP2012}) produces a similar expression. 
\end{remark}

\paragraph{Overview of techniques} The main obstacle in obtaining threshold lattice-based signatures is either rejection sampling or trapdoor sampling. They can be replaced with noise flooding (Section \ref{noiseFlooding}), in which case a threshold scheme can be  constructed following a na\"ive Schnorr-like approach based on Shamir's secret sharing (Section \ref{SecretSharing}). In the FSwA framework, it is also possible to keep the rejection sampling step, by using homomorphic commitments (Section \ref{HomomorphicCommit}). Another problem that arises in the lattice setting is how to guarantee the shortness of partial responses, to prevent leak of secret shares. This is solved by either masking partial responses or proving their shortness, when Shamir's secret sharing is used (as discussed in Section \ref{SecretSharing}); alternatively, sharing schemes with small coefficients can be used (Section \ref{alternativeSS}). Furthermore, there are other ad-hoc solutions (Section \ref{other}). Generic approaches, such as FHE and MPC, have also been employed in this paradigm and are covered in Section \ref{genericApproaches}.

\subsection{Noise Flooding} \label{noiseFlooding}
A technique often used in place of \textit{trapdoor sampling} or \textit{rejection sampling} is \textit{noise flooding}~\cite{GoldwasserKPV10}, that consists of employing a large noise to hide the secret. In short, the error term $\mathbf{r}$ in the response $\mathbf{z}=\mathbf{r}+c\cdot \mathbf{s}$ must be sufficiently dispersed. This either changes how trapdoor sampling is done, making it threshold-friendly, or allows for the removal of the rejection sampling step. 
The resulting protocols are no longer zero-knowledge and the leakage of the key must be mitigated somehow. 

There are two possible approaches for this goal: a \emph{statistical} approach (e.g.,~\cite{agrawal2022,CTZ2025,GKS2024,EKT2024}) that uses a measure---such as Renyi divergence---to determine bounds on the parameters in order to minimize the leakage; or a \emph{computational} approach (e.g.,~\cite{Pelican2024,threshRaccoon2024,Ringtail2025,KRT2024,dPKN+2025,dPEN+2025}) that reduces the security to the hardness of a variant assumption---such as the Hint-LWE problem~\cite{KLSS2023}.
In both cases, the minimization of the leakage results in bigger signatures. This technique is used in the GPV framework for Plover~\cite{plover2024} and its threshold version Pelican~\cite{Pelican2024}, and it is used in the Fiat-Shamir framework for Raccoon~\cite{Raccoon2024}, Threshold Raccoon~\cite{threshRaccoon2024}, its variants (e.g.,~\cite{EKT2024,Ringtail2025,KRT2024,dPKN+2025,dPEN+2025}) and other works (e.g.,~\cite{agrawal2022,GKS2024}).

\subsection{Na\"ive Approach with Secret Sharing} \label{SecretSharing}
Shamir's secret sharing is a fundamental tool for distributing a secret key. As in threshold Schnorr signatures, we can break up the secret key in shares, which can be used to jointly produce a signature without revealing the secret, thanks to the linearity of the underlying scheme. In short, participants are given partial secrets $\{\mathbf{s}_i\}_{i=1}^N$ such that, for every subset $S \subseteq \{1, \dots, N\}$ of cardinality $T$: $\mathbf{s}=\sum_{i \in S}\lambda_{i}^{(S)} \cdot\mathbf{s}_i$, where $\lambda_i^{(S)}$ are the Lagrange coefficients corresponding to the set $S$. To generate a signature, each participant can locally generate a partial random noise $\mathbf{r}_i$ and output a partial response $\mathbf{z}_i:=\mathbf{r}_i+ c\cdot \lambda_i\mathbf{s}_i$. The aggregated response is $\mathbf{z}:= \sum_{i \in S}\mathbf{z}_i$.  Since the resulting signature has the same formal expression as the base signature, the verification algorithm is unchanged (\textit{functional interchangeability)}. This template is followed by Threshold Raccoon~\cite{threshRaccoon2024} (and its variants). 
Pelican~\cite{Pelican2024} follows the same idea, but with a slight difference, because they obtain a sharing of $\mathbf{z}$ from the sharings of $\mathbf{s}$ and $\mathbf{r}$. As a consequence, partial responses have the same form of the base signature $\mathbf{z}_i=\mathbf{r}_i+c \cdot \mathbf{s}_i$ and the signature is aggregated through Lagrange coefficients as $\mathbf{z}=\sum_{i \in S} \lambda_i\mathbf{z}_i$.

Using noise flooding, steps that are hard to convert (such as rejection sampling) are removed, and the resulting scheme is correct thanks to the linearity of the signature. However, we still require the shortness of the partial responses to avoid leakage of partial secrets. Two possible ways to deal with it are \textit{additive masks} or \textit{verifiable short secret sharing}. Another possibility is to avoid Shamir's secret sharing and use sharing schemes with short reconstruction coefficients (see Section~\ref{alternativeSS}).

\paragraph{One-time Additive Masks} A problem with using Shamir's secret sharing is that Lagrange coefficients $\lambda_i$ can be arbitrarily large, leading to a large partial response and, thus, an insecure signature. Indeed, an attack is possible in this setting~\cite{threshRaccoon2024}, by collecting enough partial signatures from a user $i$ for different sets $S$, then solving a linear system to recover $\mathbf{s}_i$. In Threshold Raccoon~\cite{threshRaccoon2024} and some of its derivatives~\cite{EKT2024,Ringtail2025,KRT2024,dPKN+2025}, \textit{one-time additive masks} are used to hide secret shares in the partial response. The technique works as follows: (i) each pair of users $i,j \in S$ shares two random private masks $(\mathbf{m}_{i,j}, \mathbf{m}_{j,i})$; (ii) each user publishes the \textit{row mask} $\mathbf{m}_i = \sum_{j \in S} \mathbf{m}_{i,j}$; (iii) each user computes a private \textit{column mask} $\mathbf{m}^*_i = \sum_{j \in S} \mathbf{m}_{j,i}$ and adds it to their partial response as $\mathbf{z}'_i=\mathbf{z}_i+\mathbf{m}^*_i$; (iv) the final response is $\mathbf{z}= \sum_{i \in S}(\mathbf{z}'_i-\mathbf{m}_i)$, which removes the sum of column masks in the aggregation. While this also enables achieving adaptive security~\cite{KRT2024}, it makes it harder to detect misbehavers; indeed, del Pino et al.~\cite{dPKN+2025} resort to commitments and ZKPs to achieve \textit{identifiable aborts}.

\paragraph{Verifiable Short Secret Sharing (V3S)} A \textit{verifiable secret sharing (VSS) scheme} allows participants to verify that the shares received from the dealer are correct. Pelican~\cite{Pelican2024} resorts to a VSS scheme that additionally enables to verify that the shared vector $\mathbf{x}$  is \textit{short}. In the centralized version, a trusted dealer: (i) samples a value $\mathbf{y}$ from an appropriate distribution; (ii) generates and distributes shares $\mathbf{x}_i$ and $\mathbf{y}_i$, such that $\sum_{i \in S}\lambda_i\mathbf{x}_i= \mathbf{x}$ and $\sum_{i \in S}\lambda_i\mathbf{y}_i= \mathbf{y}$; (iii) hashes the shares into a Merkle tree of root $h$ and sends authentication paths $\pi_i$; (iv) computes a challenge matrix $\mathbf{R}$ (\textit{submersion matrix}), derived from $h$ using a suitable distribution, and computes $\mathbf{v}_i:=\mathbf{R} \, \mathbf{x}_i + \mathbf{y}_i$; and (v) broadcasts $h$ and the shares $\mathbf{v}_i$. Each party can verify the correctness of $\mathbf{v}_i$, using $\mathbf{R}$, $\mathbf{x}_i$ and $\mathbf{y}_i$, then they can compute $\mathbf{v}= \sum_{i \in S}\lambda_i\mathbf{v}_i$ and check that it is short. The matrix $\mathbf{R}$ must be constructed in such a way that: $\mathbf{v}$ is small if $\mathbf{x}$ and $\mathbf{y}$ are small; $\mathbf{v}$ is big if $\mathbf{x}$ is big.
This protocol is not exactly zero-knowledge because $\mathbf{v}=\mathbf{R} \, \mathbf{x} + \mathbf{y}$ leaks information on $\mathbf{x}$, but amplifying the noise $\mathbf{y}$ mitigates this leakage (\textit{noise flooding}). In Pelican, they are able to construct a 3-round distributed key generation (DKG) protocol and a 4-round signature protocol, by employing this scheme in a distributed way. Indeed, they obtain a sharing of $\mathbf{z}=\mathbf{r}+c\cdot \mathbf{s}$ from those of $\mathbf{r}$ and $\mathbf{s}$, using linearity, guaranteeing the shortness of the response and bypassing the problem on partial responses entirely.

\subsection{Homomorphic Commitments} \label{HomomorphicCommit}
In the Fiat-Shamir paradigm, we can avoid removing rejection sampling using \textit{additively homomorphic commitments}.
Usually, each party shares a \textit{commit message}\footnote{We use this name to avoid confusion with the \textit{commitment} of $\mathbf{w}_i$.} $\mathbf{w}_i$ of the random element $\mathbf{r}_i$, then everyone can compute the challenge $c= H(m, \mathbf{w})$ from  $\mathbf{w}=\sum_{i \in S}\mathbf{w}_i$. However, all $\mathbf{w}_i$ are disclosed whether the algorithm aborts or not, in which case the scheme cannot be proved secure, since information may be
leaked by aborted executions~\cite{DOTT2022,GKS2024}, which we can avoid with commitments. 

A commitment scheme is \textit{additively homomorphic} if, whenever $\mathsf{com_1}$ and $\mathsf{com_2}$ are commitments of $m_1$ and $m_2$, the message $m_1+m_2$ is a correct opening for $\mathsf{com_1}+\mathsf{com_2}$. This property essentially allows us to replace $\mathbf{w}_i$ with commitments $\mathsf{com_i}$ and replace the sum $\sum_{i\in S}\mathbf{w}_i$ with $\sum_{i \in S} \mathsf{com_i}$, so that $\mathbf{w}_i$ are not disclosed in case of abort. Since $\sum_{i \in S} \mathsf{com_i}$ is required for verification, the verification algorithm is slightly changed and, thus, functional interchangeability is lost.

A multi-signature ($N$-out-of-$N$) is constructed with this technique~\cite{DOTT2022} using Baum et al.'s homomorphic commitments~\cite{BDL+2018}, but it is not trivial to transform it into a threshold signature. Although in principle Shamir's secret sharing could be used, the number of expected repetitions due to aborts would grow too much~\cite{TPCZ2023}, because of arbitrarily long responses. Nevertheless, Tang et al.~\cite{TPCZ2023} are able to construct a threshold signature with rejection sampling, using additively homomorphic commitments and MPC based on Shamir's secret sharing (see Section \ref{genericApproaches}).
\begin{remark}
    Homomorphic commitments are used in threshold schemes also for uses other than simply hide information on aborts, such as preventing manipulation by an active adversary. In~\cite{GKS2024}, even if noise flooding is used to prevent aborts, the authors still employ Baum et al.'s commitments plus non-interactive ZKPs to make their scheme secure against an active adversary.
\end{remark}

\subsection{Alternative Secret Sharing Schemes} \label{alternativeSS}
Shamir's secret sharing is often the natural choice for threshold constructions due to its linearity and its one-share-per-participant simplicity. However, its use of arbitrary reconstruction coefficients can become problematic in lattice-based settings. An alternative is to adopt secret sharing schemes with \emph{small} reconstruction coefficients, trading increased share size for improved compatibility with lattice primitives.

\textit{Replicated secret sharing} is employed by two derivative works of Threshold Raccoon, i.e.,~\cite{dPEN+2025,PN2025}, to obtain a sharing of the secret key with small reconstruction coefficients \emph{and} short shares. In the first work, this is used with noise flooding, while in the second one with rejection sampling to obtain more compact signatures. 
We recall that, with replicated sharing, the secret is $\mathbf{s}=\sum_{j=1}^M\mathbf{s}_{I_j}$ (ranging over the class of sets such that $|I_j|=N-T+1$) and each party has shares $\mathbf{s}_i:=\{\mathbf{s}_{I_j}\}_{I_j \ni i}$ corresponding to sets they belong to. There is also a recovery function $F(S,i)$ that from a set of signers $S$ and party $i$ determines which shares $i$ must provide during reconstruction. So, partial responses will have form $\mathbf{z}_i:=\mathbf{r}_i+c\sum_{j\in F(S,i)}\mathbf{s}_{I_j}$ and they can be aggregated as a sum. This method does not require masking and, thus, it allows for \textit{partial verification} of the $i$-th signature to detect possible misbehaviours (\textit{identifiable aborts}). 
Replicated secret sharing suffers from high communication costs because the number of shares is $\binom{N}{T-1}$, which grows very quickly; therefore, it is only employed for small parameters $T$ and $N$.
In~\cite{dPEN+2025}, a \textit{ramp secret sharing} is used for bigger parameters. Ramp secret sharing has two thresholds $T_1$ (\textit{security threshold}) and $T_2>T_1$ (\textit{reconstruction threshold}), and splits the secret additively $\mathbf{s}=\mathbf{s}_1+\dots+\mathbf{s}_M$ (but with $M$ much lower than $\binom{N}{T-1}$). The idea from del Pino et al.~\cite{dPEN+2025} is to use a pseudorandom function from $\{1, \dots, N\}$ to $\{1, \dots, M\}$ to map each party to its shares, trading efficiency for a security gap between the two thresholds.

Another Threshold Raccoon variant, called Hermine~\cite{Hermine}, employs a \textit{everywhere-short secret sharing} (built on multiplicative non-abelian secret sharing~\cite{multiplicativeNonAbelianSS}), i.e. a sharing scheme that splits a short vector into short shares, while also allowing to reconstruct the secret by simply summing the shares. This enables Hermine's authors to bypass masking and obtain a lattice-based scheme most resembling FROST: it is \textit{partially non-interactive} (one round plus a preprocessing round) and it achieves non-interactive identifiable aborts. This sharing scheme remains practical for medium thresholds ($T\leq N\leq 64$). Both replicated secret sharing and multiplicative non-abelian secret sharing are also employed in the context of group actions (see Section~\ref{groupbased}).

Other linear schemes include $\{0,1\}$-$\mathsf{LSSS}$~\cite{Boneh2018,EY2024}, a class of schemes with binary coefficients, and $\mathsf{TreeSSS}$~\cite{improvedUT2025}, which has coefficients bound by a polynomial function of $N$ but has a lower number of shares, both used with threshold FHE. Finally, Benaloh and Leichter~\cite{Benaloh1990}, a scheme with coefficients in $\{-1,0,1\}$, is employed in a signature scheme~\cite{CTZ2025}.

\subsection{Other Techniques} \label{other}
\paragraph{Linear Hash Functions}
Signatures in the FSwA paradigm are obtained leveraging the similarities with Schnorr signatures with some caveats (see Section \ref{SecretSharing}). Some works (i.e.,~\cite{EKT2024,Ringtail2025}) use a trick originated in FROST~\cite{KG2020}, where the random nonces are linearly generated from a set of message-independent random elements, allowing for preprocessing and therefore removing one round from the protocol. Their security is based on variant assumptions, but a recent work~\cite{CTZ2025} is able to achieve the same from the standard SIS assumption. To do this, the authors borrow the following idea from a variant of FROST~\cite{TZ2023}: if the map $s \mapsto g^s$ in FROST is replaced by a \textit{linear hash function}, the signature security can be reduced to the standard DL assumption. 

A \textit{linear hash function (LHF)} is a linear map $F$ between two vector spaces that is also a compression function and collision-resistant. Using a LHF $F$ and two hash functions $H_1, H_2$, a 2-round threshold signature scheme can be constructed as follows: (i) each party $i$ samples $r_{i,0}$ and $r_{i,1}$ and publishes $(R_{i,0},R_{i_1}):=(F(r_{i,0}),F(r_{i,1}))$; (ii) each party $i$ sends $R:= \sum_{i \in S}R_{i,0}+b\, R_{i,1}$, with $b:=H_1(\mathsf{pk},m,S, \{R_{i,0},R_{i,1}\}_{i\in S})$, and $z_i:=r_{i,0}+c\,\lambda_i^S\, \mathsf{sk}_i$, where $c:=H_2(\mathsf{pk},m,R)$; (iii) an aggregator computes $z:=\sum_{i \in S}z_i$ and outputs $\sigma:=(R,z)$.

A natural instantiation of $F$ for lattices is $F(\mathbf{x})=\mathbf{A}\, \mathbf{x}$, which yields a scheme based on SIS. However, since the domain of $F$ is not a vector space, it requires a few adjustments for the proof of security to work and it needs the reconstruction coefficients of the sharing scheme to be small (see~\cite{CTZ2025} for more details).

\paragraph{Partial Lattice Trapdoors}
Threshold signatures in the GPV framework can also be obtained in the following way~\cite{Albrecht2025}: the full gadget trapdoor is split in partial trapdoors that are used to find partial pre-images of the target vector $\mathbf{u}$. Then, the partial pre-images are added together as usual.

The idea behind this approach is to construct the gadget matrix $\mathbf{G}$ as a block matrix with $N$ repeated blocks $\mathbf{G}_n$ on the diagonal, and split the trapdoor in blocks $\mathbf{T}_i$ corresponding to the gadget block. The target vector is also split in blocks so that each partial pre-image $\mathbf{z}_i$ can be found using $\mathbf{T}_i$. Given matrix $\bar{\mathbf{A}}\in\mathbb{Z}_q^{nN \times mN}$, partial trapdoors are $\mathbf{T}_i\in \mathbb{Z}_q^{mN\times m}$ such that $\bar{\mathbf{A}}\, [\mathbf{T}_1\,\dots\,\mathbf{T}_N]=\mathbf{I}_N\otimes\mathbf{G}_n$.
Then, the $i$-th trapdoor has gadget $\mathbf{e}_i\otimes\mathbf{G}_n$ (where $\mathbf{e}_i$ is the $i$-th canonical vector in $\mathbb{Z}_q^N$), because $\bar{\mathbf{A}}\, \mathbf{T}_i = \mathbf{e}_i\otimes\mathbf{G}_n$. To sample a pre-image of a target $\mathbf{u}$, the vector is split as $\mathbf{u}^T=\begin{bmatrix}
    \mathbf{u}_1  \cdots \mathbf{u}_N
\end{bmatrix}$ and each party can sample locally a partial pre-image $\mathbf{z}_i$ such that $\bar{\mathbf{A}}\, \mathbf{z}_i=\mathbf{e}_i \otimes\mathbf{u}_i$
%
%
This procedure only works for the $N$-out-of-$N$ case, but it can be extended to $T$-out-of-$N$, using a compressed public matrix $\mathbf{A}:=(\mathbf{V}\otimes\mathbf{I}_n)\, \bar{\mathbf{A}}$, with $\mathbf{V}$ as the Vandermonde matrix (full details in~\cite{Albrecht2025}).
The above procedure is purely algebraic and does not require use of heavy tools such as MPC or FHE, and it achieves a non-interactive threshold scheme with \textit{identifiable aborts} because partial pre-images are computed independently and are publicly verifiable. However, signatures and public keys are, respectively, linear and quadratic in size w.r.t. the threshold $T$, making it suitable only for small thresholds. Moreover, the security proof has to rely on variant assumptions.

\section{Signatures from Other Paradigms} \label{otherParadigms}

\subsection{Hash-based Signatures}
Hash-based signatures (HBS) date back to Lamport signatures~\cite{lamport1979}. The idea is to commit secret values through a \textit{one-way function}, and then reveal them to sign. The resulting scheme is a \textit{one-time signature (OTS)} since these values can only be used once. Winternitz signatures~\cite{Merkle1990} improve the scheme by deriving secret values through repeatedly computing a one-way function, instead of storing them all. Finally, \textit{Merkle tree signatures} (e.g.,~\cite{Merkle1990,XMSS2011,rfc8554}) generate $2^h$ key pairs as leaves of a Merkle tree, producing an authentication path as part of the signature, and allowing $2^h$ signatures without changing the public key, provided each leaf is used once. In Section \ref{genericApproaches}, we will describe a more general approach based on STARKs, but here we focus on a method specifically tailored to stateful HBS.

\paragraph{Threshold Signatures from Stateful HBS} \label{sigFromXORShares}
Kelsey et al.~\cite{statefulHBS} propose a framework to construct threshold signatures from stateful hash-based signatures, such as LMS~\cite{rfc8554} or XMSS~\cite{XMSS2011}. All it requires is a Merkle signature based on an underlying secure OTS. The resulting signature satisfies functional interchangeability. We first review their idea in the $N$-out-of-$N$ case.

Given a signature $(r,z,\mathsf{path})$ and its corresponding secret values $\mathsf{sk_{i,j}}$, each of these elements is split into XOR sums and each party uses those shares to produce signature shares. A \textit{trusted dealer} is required in the setup phase because they know the original one-time signature and one-time signing key, which they must split before deleting secret values. They generate one share for each party, plus a common reference value $\mathsf{CRV}$ that is used for reconstruction: for an element $x$ this means $x = \mathsf{CRV}.x \oplus \Big( \bigoplus_{s=1}^Nx^s \Big)$. The values in $\mathsf{CRV}$ must be stored, in order to be later accessed by an untrusted \textit{aggregator}, that interacts with signers to produce the signature. One of the signers could be the aggregator without undermining the security of the scheme. The signing process is the following: (i) each party uses their key $k_s$ to derive shares\footnote{$\mathsf{chk}$ is a vector of values required to check integrity of $r$.} $r^s$ and $\mathsf{chk}^s$ and sends them to the aggregator; (ii) the aggregator uses the $\mathsf{CRV}$ share and each party's share to compute $r$ and $\mathsf{chk}$ and sends them back; (iii) each party checks that $r$ generates the $s$-th element of $\mathsf{chk}$ through a pseudorandom function; then, if the check is valid, each party uses their key $k_s$ to derive $z^s$ and $\mathsf{path}^s$ and sends them back; (iv) the aggregator computes $z$ and $\mathsf{path}$ from the shares and $\mathsf{CRV}$; then, outputs $(r,z,\mathsf{path})$.

This scheme can be turned into a $T$-out-of-$N$ signature simply adding a step in which the aggregator determines a coalition of $T$ signers and follows the protocol for a $T$-out-of-$T$ signature. However, it is important for security that no key is used twice. Therefore, Kelsey et al.\cite{statefulHBS} propose \textit{sharding} as a solution. Each OTS is associated with a unique identifier $\mathsf{KeyID}$ that is used once. For simplicity, we may assume that $\mathsf{KeyID} \in \{1,\dots, D\}$ and we can break this set into about $D/\binom{N}{T}$ ranges (called \textit{shards}). Each shard is associated only with one coalition, so that two coalitions never sign with the same identifier. For this purpose, during the setup, the dealer distributes to each party the coalitions to which they belong.

Observe that this framework necessarily has a centralized setup, and it is not feasible for stateless HBS due to the massive number of shares to store in the $\mathsf{CRV}$. Moreover, the number of coalitions grows as $\binom{N}{T}$.

\subsection{Based on Group Actions}
\label{groupbased}
Schemes based on group actions leverage the hardness of the \textit{Group Action Inverse Problem (GAIP)}, i.e., for a group action $\star:G \times X \to X$, given $x,y \in X$, it should be hard to compute $g \in G$ such that $y = g \star x$. This can be seen as a generalization of the DLP, and therefore these signatures are built using an identification scheme similar to Schnorr's~\cite{Schnorr1991}, then applying the Fiat-Shamir transform~\cite{FS1987}: (i) a random $h\in G$ is sampled and $h \star x_0$ is computed from public $x_0 \in X$; (ii) a challenge is derived as $c=H(h\star x_0, m) \in \{0,1\}$; (iii) the response is $(c,\mathsf{rsp})$ with $\mathsf{rsp}=hg^{-c}$. If we assume the GAIP is hard, the soundness error is $1/2$, so we will have to repeat it $\lambda$ times to decrease it to $1/2^{\lambda}$.

\paragraph{Threshold Evaluation of the Group Action} First, we review how cyclic group actions can be evaluated in a distributed way. Then, we consider general (possibly non-abelian) groups.

In case of cyclic group action, we use notation from \textit{isogeny-based cryptography} with $[s]E$ for $g^s\star E$. To evaluate the action of a sum $[\sum_ir_i]E^0$ on a public element $E^0$, where each element $r_i$ is secret and owned by a party, the parties have to decide on an order of computation between them and adopt a sequential \textit{round-robin} communication structure. The first party computes $E^1:= [r_1]E^0$ and then sends the output to the second party. At the $k$-th round of the protocol, the $k$-th party, upon receiving $E^{k-1}$ computes $E^k:=[r_k]E^{k-1}$. Since $E^k=[r_k]E^{k-1}=[r_k][r_1+\dots+r_{k-1}]E^0=[r_1+\dots+r_k]E^0$, by induction, the final result is $[\sum_ir_i]E^0$. Since each step involves the computation of a group action, the secret $r_i$ stays secret thanks to the GAIP. As long as we employ linear secret schemes, we can split the secret key and compute $[\sum_{i \in S}\lambda_is_i]E^0=[s]E^0$ to obtain the public key. We remark that this computation is inherently \emph{sequential}, and therefore it leads to a high number of rounds for the protocol. Moreover, existing isogeny-based implementations, e.g.,~\cite{HHS2006,CSIfish2019,DeFeoMeyer2020}, are computationally intensive, although they benefit from small keys and small signatures.

The evaluation of the group action for a general (possibly non-abelian) group is more complicated in a distributed setting. Battagliola et al.~\cite{GRASS} do it in the full threshold case ($T=N$) and then extend it to the threshold setting using replicated secret sharing (see Section \ref{alternativeSS}), since there is no obvious way to use Shamir's in a non-abelian setting; therefore, it is practical only for small $N$. However, this method is more general and effectively also generalizes \emph{code-based schemes}, such as LESS~\cite{LESS2020} or MEDS~\cite{MEDS2023}, that require non-abelian actions. For $T=N$, the action $g\star x_0:=(g_M\dots g_1) \star x_0$ is evaluated sequentially as $x_k:=g_k \star x_{k-1}$ and the $g_i$ must be applied in a specific order.

\paragraph{Threshold Signatures from Group Actions} In the cyclic case, a threshold signature can be obtained as follows: (i) during their round $k$, the party $P_i$ uniformly samples $r_{i,j} \in \mathbb{Z}_q$ for $j=1,\dots,\lambda$ and computes $E^k_j=[r_{i,j}]E^{k-1}_j$. The above protocol is used to derive $E^T_j:=[\sum_{i \in S}r_{i,j}]E^0$; (ii) each party computes $(c_1,\dots,c_{\lambda})=H(E^T_1,\dots E^T_{\lambda}, m)$, where $H:\{0,1\}^*\to \{0,1\}^{\lambda}$ is a hash function; (iii) each party outputs $z_{i,j}=r_{i,j}-c_j\cdot \lambda_is_i$. The final response is $(c_1,\dots,c_{\lambda},z_1,\dots,z_{\lambda})$ where $z_j = \sum_{i \in S}z_{i,j}$. Verification is done as the base signature (\emph{functional interchangeability)}.
In this way, a threshold version of CSI-FiSh~\cite{CSIfish2019} with a trusted dealer is obtained~\cite{DeFeoMeyer2020}. Follow-up works improve it in the following ways: they employ a DKG to remove the trusted dealer and they are able to achieve active security~\cite{sashimi2020,CSIrashi2021,CamposMuth2022,CSIshark2023}; and, using larger challenge spaces, they reduce the number of iterations $\lambda$~\cite{sashimi2020} (since this involves sharing many secrets, this increases computation and communication costs for DKG; however, this can be mitigated with structured public keys~\cite{CSIshark2023}).

For the non-abelian setting, the procedure is similar, but the response is also computed sequentially as $\mathsf{rsp}_{k}:= h_{k-1}\mathsf{rsp}_{k-1}g_{k-1}^{-c}$. The GRASS framework~\cite{GRASS} constructs threshold signatures from generic group actions this way, assuming that the GAIP is hard and that the group action satisfies a weaker notion of pseudorandomness (\textit{2-weakly pseudorandom}). A later refinement, GRASS+~\cite{grass+}, improves efficiency by replacing replicated secret sharing with multiplicative non-abelian secret sharing inspired by the work of Desmedt et al.~\cite{multiplicativeNonAbelianSS}, and achieves adaptive security reducing it to a variant of the GAIP.

\section{Generic Approaches} 
\label{genericApproaches} Besides the methods specifically tailored to some paradigm, there are also general approaches. We briefly recall here the main ideas and how they have been applied.

\paragraph{Threshold Fully Homomorphic Encryption} TFHE offers a black-box framework for constructing threshold signatures from signatures. The idea is to homomorphically evaluate the signing circuit, then jointly decrypt to output the signature. Boneh et al.~\cite{Boneh2018} build a \textit{universal thresholdizer (UT)} from \textit{threshold fully homomorphic encryption}, that turns a signature scheme into a one-round threshold signature, later improved to also provide active security~\cite{EY2024,agrawal2022}. In principle, the UT can be used with signatures from any paradigm. However, existing FHE schemes are almost all based on lattice techniques~\cite{FHEsurvey,FHEguide}, and UT has only been instantiated on a lattice-based signature~\cite{agrawal2022}, using noise flooding (Section \ref{noiseFlooding}) to remove the rejection sampling step, since it is hard to evaluate homomorphically. We remark that this approach likely incurs computation and communication overhead due to FHE evaluation of the circuit, and requires to adjust the underlying signature to be more amenable to evaluation. However, it minimizes the number of rounds of the protocol (to only one) and it also satisfies functional interchangeability.

On the other hand, Gur et al.~\cite{GKS2024} introduce a framework, based on \textit{linearly homomorphic encryption}, that does not require a \emph{fully} HE scheme, but it is only applicable to signatures made of linear operations. They construct a threshold version of the BGV encryption scheme~\cite{BGV2012} and use it to build a threshold signature from a variant of Dilithium~\cite{dilithium2018} that replaces rejection sampling with noise flooding. Since the response $\mathbf{z}$ only requires additions and multiplications by a public scalar, using the linearly homomorphic scheme is enough. While this is a less general solution than Boneh et al.'s UT, it produces a more efficient scheme by removing parts that are expensive to evaluate with HE.

\paragraph{Multi-Party Computation}
MPC can be used to jointly evaluate the signing algorithm of an underlying signature scheme, by composing MPC core functionalities, such as multiplication, with other protocols, such as comparison protocols. In principle, one could apply generic MPC protocols in a black-box manner to existing signatures. However, this approach rarely produces practical results~\cite{CozzoSmart19}, especially for non-linear operations, because they require conversion from MPC based on linear secret sharing to MPC based on garbled circuits. Moreover, many generic protocols are meant for the full threshold case ($T=N$) and have to be adapted for the $T$-out-of-$N$ case~\cite{TPCZ2023}.
A big advantage of using MPC is to benefit from its properties: many protocols are constructed with security against active adversaries in mind, that must be otherwise achieved with other techniques such as commitments and non-interactive ZKPs~\cite{GKS2024}. Moreover, since the underlying signature is evaluated directly, the threshold signature satisfies functional interchangeability. However, it requires a lot of one-time pre-processed triples for multiplications~\cite{Beaver1992}.

Tang et al.~\cite{TPCZ2023} and Bienstock et al.~\cite{Bienstock2025} achieve a threshold signature secure against active adversaries in the FSwA setting, carefully adapting parts of the scheme that would be difficult to compute. Both realize rejection sampling using MPC. On the other hand, Bendlin et al.~\cite{BKP2013} use MPC functionalities to evaluate a GPV signature based on gadget trapdoors~\cite{MP2012}. We also remark that Bienstock et al. and Bendlin et al. work within the \textit{Universal Composability (UC)} framework~\cite{UCframework2001}, meaning their protocol stays secure even if used as a component of a larger protocol; instead, Tang et al. only prove unforgeability in the traditional sense.
The only signature schemes not based on lattices yet approached with MPC are those based on multivariate quadratic (MQ) problems, and in particular OV-based, that follow a hash-and-sign approach: they require the inversion of a trapdoor function, done by sampling a random secret vector and solving a linear system. Therefore, they are amenable to evaluation through MPC. This has been investigated by Cozzo et al.~\cite{CozzoSmart19} with good results for LUOV~\cite{LUOV2017} and Rainbow~\cite{rainbow}, and a threshold version of MAYO~\cite{Mayo2022} has been proposed following a similar approach~\cite{thresholdMayo2025}. However, OV-based schemes have a long history of attacks, and in particular LUOV and Rainbow have been broken for some parameter sets (e.g., \cite{variantUOVattacks2006,UOVattacks2009,LUOVattacks2020,UOVRainbowAttacks2019,UOVRainbowAttacks2021,rainbowAttacks2022,MAYOattack2025}).

\paragraph{STARKs} Informally, STARKs~\cite{STARKs2018} are ZKPs with the properties of being \textit{scalable}---proving and verification times do not grow too much---and \textit{transparent}---not requiring a trusted setup phase. They are also PQ secure. Typically, they are interactive proofs, but they can be converted to non-interactive in the random oracle model, using Fiat-Shamir. This transformation is secure even in the quantum setting when the underlying proof-of-knowledge is secure~\cite{FiatShamirQROM_Don19,FiatShamirQROM_Liu19}. At a high level, to construct a threshold signature scheme, we require a setup phase to commit to a set of $N$ public keys, using an \textit{accumulator}\footnote{Accumulators are cryptographic primitives that allow to prove membership of an element to a set.}, such as a Merkle tree. Since each party independently generates keys and signatures, no DKG protocol is required. Then, during the signing phase, an aggregator collects signatures and proves that they are valid and they are at least $T$, therefore obtaining a one-round scheme. This approach is followed only by one hash-based signature~\cite{Khaburzaniya2022}, but in principle it could be used in any paradigm. The main drawback is that it requires an intermediate step, the \textit{Algebraic Intermediate Representation}, to represent the signature verification algorithm with polynomial statements. This process can be arbitrarily difficult and can lead to inefficiency. Indeed, Khaburzaniya et al.~\cite{Khaburzaniya2022} have to replace standard hash functions with an ad-hoc ZKP-friendly one. Moreover, the verification of the threshold signature consists of verifying ZK statements. This has two consequences: functional interchangeability is lost, and the signature size grows with the threshold $T$, albeit sub-linearly thanks to the compactness of STARKs.

\section{Conclusion}

The techniques considered in this work are not exclusive to the PQ setting, but leverage existing ideas to build threshold schemes. Most of the originality of these works is to adapt existing tools to specific settings. The main idea is to use a linear secret sharing scheme and aggregate partial responses constructed from secret shares, as done for Schnorr signatures in FROST~\cite{KG2020}. However, PQ introduces obstacles that are not present in the classical setting. Lattice-based signatures are based on constrained versions of linear systems, and therefore have to guarantee the shortness of vectors: this introduces steps that are hard to thresholdize, and are usually tackled with noise flooding, homomorphic commitments, or ad-hoc techniques. Signatures based on group actions are a direct generalization of those based on the discrete logarithm; therefore, they can use the same ideas, but the threshold evaluation of the group action is inherently sequential, leading to protocols with many rounds. Hash-based signatures are very simple but they cannot use the same techniques as above, because they cannot be aggregated additively; therefore, it is hard to make them efficient, or they may have limitations such as requiring a trusted setup. Generic approaches (FHE, MPC, and STARKs) provide in principle black-box mechanisms to transform signatures into threshold ones. However, doing so is usually inefficient and the underlying schemes have to be adapted to be more amenable to these techniques (e.g., removing parts that are too intensive to compute with MPC). Among FHE-based approaches, UT~\cite{Boneh2018} provides a mechanism to transform a signature into a one-round threshold scheme, and has been improved to be instantiated in practical scenarios. MPC protocols often require extension to the threshold case, but provide additional properties by default, such as composable protocols and active security. MQ-based schemes seem to be the most amenable to this approach, but may raise security concerns due to significant recent attacks. STARKs cannot provide functional interchangeability and the aggregated signature size grows with the threshold, but they produce one-round signatures with no DKG required.

%
\bibliographystyle{splncs04}
\bibliography{biblio}

\end{document}